\documentclass[aip,apl,reprint,citeautoscript,superscriptaddress]{revtex4-1}

\usepackage{amsmath}
\usepackage{amsfonts}
\usepackage{amssymb}
\usepackage{graphicx}
\usepackage{color}
\usepackage{subfigure}

\newcommand{\algao}{(Al$_x$Ga$_{1{\rm -}x})_2$O$_3$}

\newcommand{\beginsupplement}{%
    \setcounter{section}{0}
    \renewcommand{\thesection}{S\Roman{section}}
    \setcounter{table}{0}
    \renewcommand{\thetable}{S\arabic{table}}%
    \setcounter{figure}{0}
    \renewcommand{\thefigure}{S\arabic{figure}}%
    \setcounter{equation}{0}
    \renewcommand{\theequation}{S\arabic{equation}}%
}

\begin{document}
\title{Origin of donor compensation in monoclinic~\algao~alloys}

\author{Sierra Seacat}
\affiliation{Department of Physics and Astronomy, University of Kansas, Lawrence, KS 66045, USA}
\email{sierra.c.seacat.ctr@us.navy.mil}

\author{Hartwin Peelaers}
\affiliation{Department of Physics and Astronomy, University of Kansas, Lawrence, KS 66045, USA}
\email{peelaers@ku.edu}

\begin{abstract}
~\algao~alloys are frequently used in heterostructures with monoclinic Ga$_2$O$_3$, resulting in a large conduction-band offset, which leads to charge carrier confinement, a property that is desirable for device applications. However, when ~\algao~ alloys are $n$-type doped with Si, the most efficient shallow donor, there is a significant reduction in the number of charge carriers when the Al content of the alloys is greater than 26\%, rendering intentional doping ineffective. Here we show that this compensation is due to cation vacancies forming in response to donor doping. We use density functional theory with the HSE06 hybrid functional to study cation vacancies in monoclinic AlGaO$_3$ and monoclinic Al$_2$O$_3$. We find that vacancies prefer to occupy split-vacancy configurations, similar to vacancies in Ga$_2$O$_3$. Furthermore, by comparing the formation energy of the vacancy with the formation energy of Si donors, we show that vacancies are lower in energy than Si donors, independent of the Fermi level, as soon as the alloys contain more than 16\% Al under O-poor conditions. Therefore, cation vacancies will compensate the donor doping, explaining experimental observations.
\end{abstract}

\maketitle

Ga$_2$O$_3$, which has a wide bandgap of 4.7 eV~\cite{Tippins1965,Matsumoto1974} and a large critical breakdown field of 5-9 MVcm$^{-1}$~\cite{Pearton2018a,Higashiwaki2017}, is among the most promising ultrawide-bandgap materials for the development of power devices, such as Schottky barrier diodes~\cite{Ji2022}, MOSFETs, and rectifiers.~\cite{Pearton2018b, Zhou2019} Additionally, the wide bandgap leads to UV transparency~\cite{Ueda1997, Orita2000}, enabling the development of solar-blind photodetectors and other optoelectronic devices.~\cite{Guo2019, Wang2021, Sun2024} 

Device fabrication often requires creating heterostructures between Ga$_2$O$_3$ and another material. Typically, these heterostructures are formed using alloys of Al$_2$O$_3$ and Ga$_2$O$_3$, which provide large conduction-band offsets with Ga$_2$O$_3$, enabling charge-carrier confinement~\cite{Ahmadi2017a, Krishnamoorthy2017, Joishi2019,Okumura2019, Chatterjee2020, Vaidya2021, Tadjer2021, Peelaers2018}, which can be used to create a two-dimensional electron gas (2DEG).~\cite{Zhang2018, Ranga2021} Monoclinic AlGaO can be synthesized through a variety of growth methods. Czochralski bulk crystals with up to 35\% Al content can be grown from the melt.~\cite{Galazka2023} Crystalline AlGaO thin films can be reliably grown using MOCVD, with films possessing up to 99\% Al content achievable using trimethylgallium (TMGa) and trimethylaluminum (TMAl) as precursors~\cite{Bhuiyan2023}. Furthermore, intentional donor doping can be achieved in these alloys, with Si being the most common and effective $n$-type dopant~\cite{Varley2020, Mu2022, Galazka2023}.

However, $n$-type doped AlGaO alloys suffer from electrical compensation, even at modest Al concentrations. Experiments have shown that the free carrier concentration decreases in bulk alloys with Al content greater than 20\%~\cite{Galazka2023}. This is far short of what is predicted by theory, as density functional theory (DFT) calculations show that Si should act as a shallow donor in alloys with up to 70\% Al content~\cite{Mu2022}. Experimentally, in Ga$_2$O$_3$/\algao~heterostructures, donor doping has only been achieved in heterostructures with up to 26\% Al content.~\cite{Ranga2020a}

One mechanism that could explain the compensation of donor doping in these alloys is the formation of cation vacancies. 
In Ga$_2$O$_3$, gallium vacancies (V$_{\text{Ga}}$) have low formation energies, so they are present in high concentrations in both bulk and thin film samples~\cite{Weiser2018, Son2020, Karjalainen2020, Johnson2019, Tuomisto2024}. Since V$_{\text{Ga}}$ acts as a deep acceptor in the 3$-$ charge state~\cite{Varley2011,Kyrtsos2017,Zimmerman2020, Frodason2021}, with the ability to bind up to three electrons, they can lead to a decrease in the charge carrier concentration. Additionally, V$_{\text{Ga}}$ has been shown to form in response to Si doping~\cite{Korhonen2015}. 
Experimentally, it is challenging to study vacancy defects. Modeling with DFT can be used to elucidate the properties of these defects, but computational studies of cation vacancies in~\algao~alloys have not yet been performed. 

Here, we address this knowledge gap by using DFT with hybrid functionals to investigate cation vacancies in monoclinic~\algao~alloys doped with Si.
We calculate the concentration of vacancy defects relative to the donors by determining the formation energies of the vacancy defects and Si dopants. These calculated formation energies allow us to determine the overall extent of vacancy compensation.  
We investigate the monoclinic structures of Ga$_2$O$_3$,  Al$_2$O$_3$ and the AlGaO$_3$ alloy, allowing us to extrapolate our results across the entire range of Al-Ga alloy concentrations. Our calculations predict that cation vacancy formation increases in alloys with at least 16\% Al content, eventually completely compensating Si doping. Furthermore, we find that the formation of vacancy defects can be limited by using growth methods that provide O-poor conditions. Note that while we focused on Si doping, which was shown to be the most efficient $n$-type donor in these alloys~\cite{Varley2020}, our conclusions will remain valid for other dopants, except that vacancy compensation will occur at lower Al concentrations. 

\begin{figure*}[tb]
	\centering
	\includegraphics[width=0.9\textwidth]{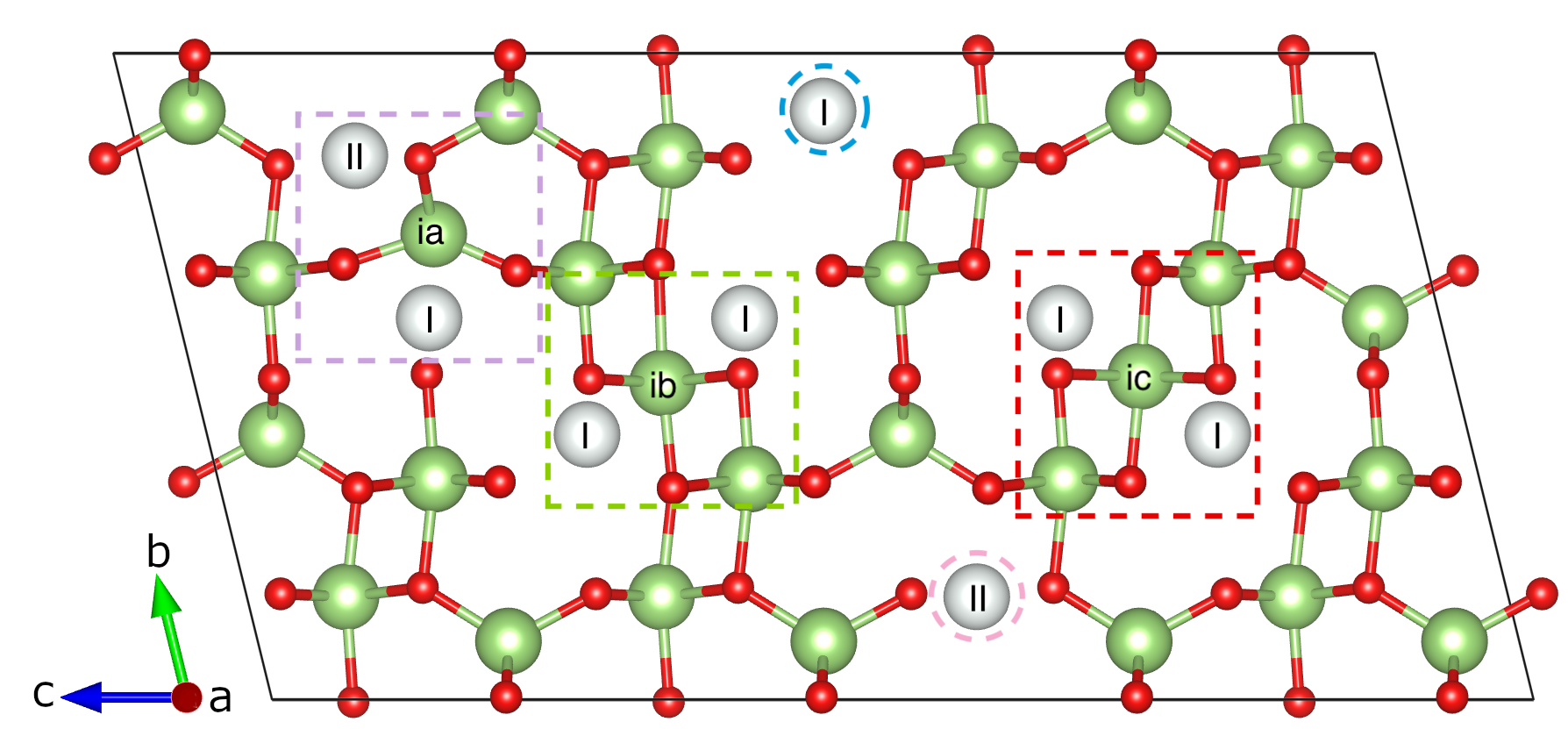}
	\caption{\label{fig:split_pos} All possible vacancy positions in monoclinic Ga$_2$O$_3$. The vacancy positions are represented by the white spheres and labeled according the coordination of the atom that was removed with I representing a tetrahedrally-coordinated cation and II representing an octahedrally-coordinated cation. The split positions are outlined by dashed boxes and the bare positions by dashed circles. The ia configuration (light violet box) involves both a tetrahedral (I) and octahedral (II) site, while the ib (bright green box) and ic (red box) involve two tetrahedrally-coordinated cations. }
\end{figure*}

All calculations were performed using the Vienna Ab-initio Simulation Package (VASP)~\cite{Kresse1993, Kresse1996} with projector augmented wave (PAW) pseudopotentials~\cite{Blochl1994}. The Ga pseudopotentials included the \textit{d}-electrons in the valence. To ensure adequate localization of the defect charge, we used the HSE06 hybrid functional~\cite{Heyd2003, *Heyd2006} with 32\% mixing. The structures containing the defects were created using 120-atom supercells made from the conventional monoclinic unit cell and were relaxed using a k-point grid consisting of a single special k-point, ($\frac{1}{4}$, $\frac{1}{4}$, $\frac{1}{4}$), and a plane-wave expansion cutoff of 400 eV until all the forces were smaller than 10 meV/\AA. Test calculations performed using the AlGaO$_3$ alloy show that the single special k-point yields a value for the crossing point between Si$_{\text{I}}$ and V$_{\text{Ga}}^{ic}$ that is within 0.6 meV of the value obtained when using a $\Gamma$-centered 2$\times$2$\times$2 k-grid. Previous calculations on defects in ~\algao~alloys have shown the 120-atom supercell to give converged results relative to the 160-atom cell.~\cite{Mu2022} 

The defect formation energy is a measure of how likely a defect is to form in a material, thus directly related to the defect concentration~\cite{Freysoldt2014}. As an example, the formation energy of a V$_{\textrm{Ga}}$ in Ga$_2$O$_3$ in charge state $q$ is

\begin{align}\label{eq:defect} 
&\text{E}^{f}(\text{Va}_{\text{Ga}}^{q}) = \text{E}_{\text{tot}}(\text{Va}_{\text{Ga}}^{q}) \rm{ - } E_{tot}(\text{Ga}_2\text{O}_3)  \notag \\ 
&\qquad  + (\mu_{\text{Ga}} + \mu^{0}_{\text{Ga}}) + q(\text{E}_{\text{F}} + \text{E}_{\text{VBM}}) +\Delta^{q} ,\ 
\end{align}

where E$_{\text{tot}}(\text{Va}_{\text{Ga}}^{q})$ is the energy of the cell containing the vacancy defect in charge state $q$, E$_{\text{tot}}$($\text{Ga}_2\text{O}_3$) is the energy of the cell without any defects, $\mu_{\text{Ga}}$ is the chemical potential for Ga referenced to the bulk value, $\mu^{0}_{\text{Ga}}$. These chemical potentials correspond to the energy required to remove the Ga atom from the system. The chemical potentials are chosen to represent theoretical limiting conditions: O-rich ($\mu_{\text{O}}=0$) and O-poor ($\mu_{\text{Ga}}=0$). For Si, the chemical potentials are determined using SiO$_2$ as the limiting phase. Details about the chemical potentials used are given in the supplemental information, so that the formation energy diagrams can be adjusted to reflect specific experimental conditions. Next, we consider Fermi levels within the bandgap, and therefore reference E$_{\text{F}}$ to the valence-band maximum (E$_{\text{VBM}}$). Finally, $\Delta^{q}$, the defect-charge correction term, is added to account for charge interactions that arise due to the periodic boundary conditions. This term is calculated using the \textsc{sxdefectalign} code~\cite{Freysoldt2011}. The dielectric tensors of monoclinic Al$_2$O$_3$ and AlGaO$_3$ were explicitly calculated using density functional perturbation theory with the PBE functional. The experimentally measured dielectric tensor was used for the calculations involving $\beta$-Ga$_2$O$_3$. 

\begin{figure*}[tb]
	\centering
	\includegraphics[width=0.9\textwidth]{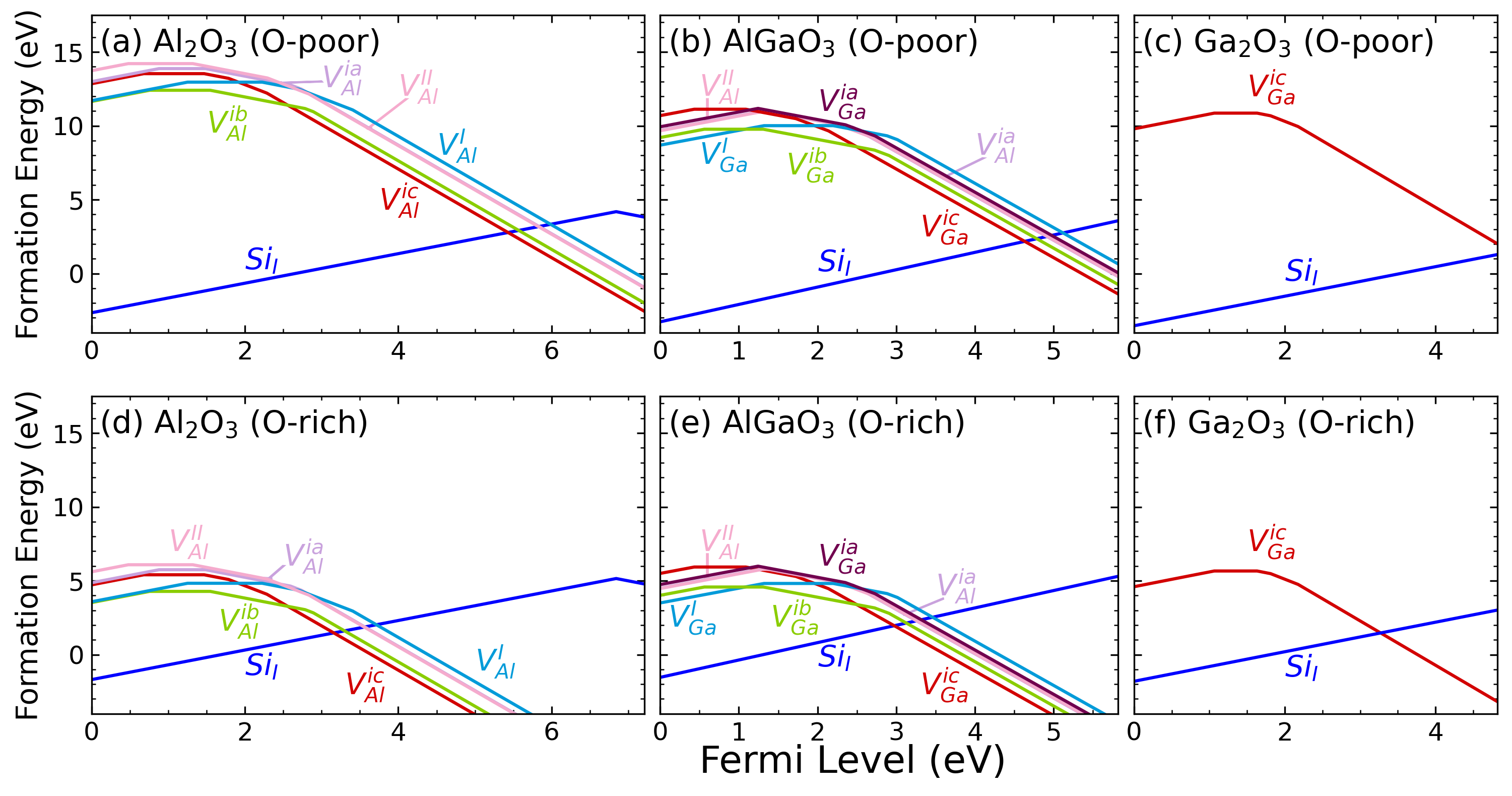}
	\caption{\label{fig:vac_defects} The formation energies for cation vacancy and Si$_\text{I}$  defects in both O-poor [(a)-(c)] and O-rich conditions [(d)-(f)] for monoclinic Al$_2$O$_3$ [(a), (f)], AlGaO$_3$ [(b), (e)], and Ga$_2$O$_3$ [(c), (f)].}
\end{figure*}

Investigating vacancy compensation in monoclinic alloys is complicated by the many different vacancy sites in $\beta$-Ga$_2$O$_3$. In addition to tetrahedral (I) and octahedral (II) bare vacancy positions, there are also three split-vacancy positions, which are lower in energy than the bare vacancies~\cite{Frodason2023, Varley2011, Tuomisto2023}. In these split-vacancy configurations, a Ga atom is removed and a second Ga atom moves to occupy an interstitial position with two half vacancies adjacent to it~\cite{Varley2011}. All five vacancy positions are given in Fig.~\ref{fig:split_pos} with the split-vacancy complexes outlined by dashed boxes and the bare vacancies outlined by dashed circles. The Ga I bare vacancy is outlined in teal and the Ga II vacancy in pink. Two of the split-vacancy configurations (ib (outlined in bright green) and ic (outlined in red)) involve two Ga I sites: one tetrahedrally-coordinated Ga is removed, while the second moves into an octahedrally-coordinated interstitial position. The ia configuration (outlined in light violet) involves a Ga I and Ga II site, with the interstitial becoming tetrahedrally-coordinated. In Ga$_2$O$_3$, the ic configuration has the lowest formation energy~\cite{Frodason2023}. However, all three split-interstitial vacancy configurations are lower in energy than the bare vacancies and have been experimentally observed~\cite{Johnson2019, Karjalainen2020, Zhelezova2024}. 

Fig.~\ref{fig:vac_defects} shows the formation energies as a function of Fermi level in oxygen-poor conditions for (a) Al$_2$O$_3$, (b) AlGaO$_3$, and (c) Ga$_2$O$_3$, and in oxygen-rich conditions for (d) Al$_2$O$_3$, (e) AlGaO$_3$, and (f) Ga$_2$O$_3$. Each vacancy configuration is depicted with a different colored line, with the slope of the line representing the charge state of the defect. Of the split configurations, V$^{\text{ia}}$ is shown in light violet, V$^{\text{ib}}$ in bright green, and V$^{\text{ic}}$ in red. In the alloys, the other V$^{\text{ia}}$ configuration with the tetrahedrally-coordinated cation removed, is represented in dark violet, while the configuration with the octahedrally-coordinated cation removed is shown in light violet. The energies for the bare vacancies, V$^{\text{I}}$  and V$^{\text{II}}$ are given by the bright blue and light pink lines, respectively. Only the lowest energy configuration is shown for Ga$_2$O$_3$. The formation energies for all possible Ga$_2$O$_3$ vacancy configurations can be found in Ref.~\onlinecite{Frodason2023}. The formation energy for a Si donor is shown using a blue line. 

In Ga$_2$O$_3$, [Fig.~\ref{fig:vac_defects}(c) and (f)], the V$^{\text{ic}}$ position has the lowest formation energy. As shown in Ref.~\onlinecite{Frodason2023}, V$^{\text{ib}}$ is slightly higher in energy compared to V$^{\text{ic}}$, while V$^{\text{ia}}$ has a formation energy comparable to V$^{\text{II}}$, and V$^{\text{I}}$ is the highest energy vacancy defect. This energetic ordering remains the same in AlGaO$_3$ [Fig.~\ref{fig:vac_defects}(b) and Fig.~\ref{fig:vac_defects}(e)], as the split vacancies are lower than the bare vacancies. V$_{\text{Ga}}^{\text{ic}}$ has the lowest formation energy, followed by V$_{\text{Ga}}^{\text{ib}}$ and both ia configurations (V$_{\text{Ga}}^{\text{ia}}$ or V$_{\text{Al}}^{\text{ia}}$), which have similar formation energies. Finally, for the bare vacancies, removing an Al II atom yields formation energies comparable to the ia vacancies. This is also more favorable compared to removing a Ga I atom, which is the highest energy vacancy defect. 
For Al$_2$O$_3$, shown by panels (a) and (d), the energetic ordering of the vacancies is the same as in Ga$_2$O$_3$ and AlGaO$_3$, with V$_{\text{Al}}^{\text{ic}}$ having the lowest formation energy. 

As defect concentrations depend on the growth temperature and the specific experimental chemical potentials, we will use the Fermi-level position at which a vacancy defect becomes lower in formation energy than the Si donor as a qualitative measure for the extent of vacancy compensation in the alloys, as its value gives an indication of when vacancy compensation will start occurring.  
This point can be calculated using 

\begin{align}\label{eq:crossing}
	\epsilon(\text{V}^{q}/\text{Si}^{q'}) = \frac{\text{E}^{f}(\text{V}^{q}; \text{E}_{\text{F}} = 0) \rm{-} E^{f}(Si^{q'}; E_{F} = 0)}{q'-q},
\end{align}

where E$^{f}$(V$^{q}$; E$_F$ = 0) is the formation energy of the lowest energy vacancy defect in charge state $q$ at the VBM and  E$^{f}$(Si$^{q'}$; E$_F$ = 0) is the formation energy of Si in charge state $q'$ at the VBM. Note that in all cases, $q = -3$ and $q'= 1$. These values are given for Al$_2$O$_3$, AlGaO$_3$, and Ga$_2$O$_3$ for both O-poor and O-rich environments in Table~\ref{tab:defects}.

\begin{figure}[tb]
	\centering
	\includegraphics[width=0.85\columnwidth]{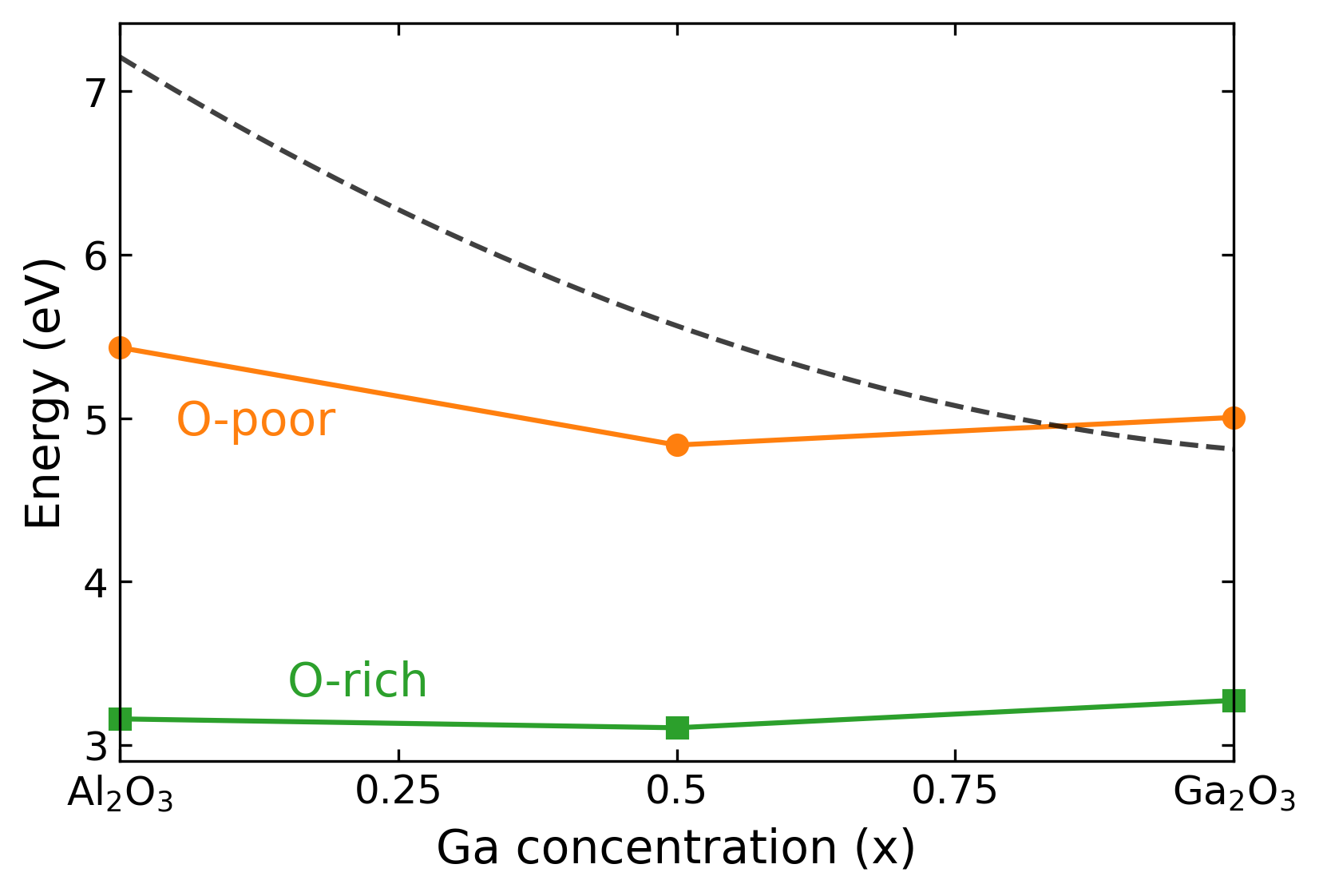}
	\caption{\label{fig:compensation} The Fermi level where the formation energy of the lowest energy cation vacancy intersects with the formation energy for the Si donor as a function of alloy composition. The solid lines are linear interpolations. The orange circles are the levels for O-poor conditions and the green squares for O-rich conditions. The black dashed line is the CBM as function of alloy concentration based on the previously determined bowing parameter~\cite{Peelaers2018}. }
\end{figure}

Fig.~\ref{fig:compensation} plots the Fermi level where the formation energy of the lowest energy vacancy defect crosses the formation energy for Si$_{\text{I}}$ [see Eq.(~\ref{eq:crossing})] as a function of alloy composition for both O-poor (orange circles) and O-rich conditions (green squares). These points are interpolated with straight lines. The level of the CBM is given by the black dashed line and was calculated using the previously determined value for the bowing parameter~\cite{Peelaers2018}. From Fig.~\ref{fig:compensation} and Table ~\ref{tab:defects}, we see that this transition point will only occur above the CBM for pure Ga$_2$O$_3$ or alloys with small Al concentrations in O-poor conditions. 

\begin{table}[tb]
	\caption{\label{tab:defects}  The bandgaps and Fermi levels (eV) where the formation energy of the vacancy defect equals the formation energy of the Si donor in Al$_2$O$_3$, AlGaO$_3$ and Ga$_2$O$_3$ for O-poor and O-rich chemical potentials. }
	\begin{ruledtabular}
		\begin{center}
	\centering
		\begin{tabular}{llll}
 	         	&Al$_2$O$_3$& AlGaO$_3$ & Ga$_2$O$_3$ \\
			\hline
			Bandgap & 7.21 & 5.82 & 4.82 \\
			\hline
			$\epsilon$(V/Si) O-poor & 5.43 & 4.84 & 5.00 \\
			$\epsilon$(V/Si) O-rich & 3.16 & 3.11 & 3.27 \\
	\end{tabular}
\end{center}
\end{ruledtabular}
\end{table}

From Fig.~\ref{fig:compensation} we can estimate that the concentration of vacancies will greatly increase once the alloys reach 16\% Al content. At this concentration, the transition point between the vacancy defect and the Si donor starts to occur within the bandgap, indicating that the formation energy of cation vacancies is lower than the formation energy of Si donors. The concentration of cation vacancies will therefore increase, and will eventually reach high enough concentrations to completely compensate intentional Si doping. The low formation energies of these vacancies compared to Si under all but the most O-poor conditions, explains why Si is only an effective donor in AlGaO alloys with Al concentrations of 26\%~\cite{Ranga2020a} in thin films or 20\% in bulk samples.~\cite{Galazka2023} Note that both growth techniques have conditions that are closer to the theoretical O-poor than the O-rich conditions, consistent with our results that O-poor conditions are optimal to allow for n-type doping in monoclinic AlGaO alloys.

Moving from O-poor to O-rich conditions will also increase the number of vacancies. O-rich growth conditions lower the formation energies for cation vacancies significantly compared to O-poor conditions. This causes the crossing point between the vacancy and the Si donor to occur well within the bandgap for Ga$_2$O$_3$ and closer to midgap for AlGaO$_3$ and Al$_2$O$_3$ (green squares in Fig.~\ref{fig:compensation}). This suggests that using O-poor growth conditions and methods to fabricate AlGaO layers is essential to limit cation vacancy compensation as much as possible. While the chemical potentials used in these calculations are theoretical limits and do not necessarily correspond to experimental conditions, the information provided in the supplemental information can be used to calculate defect formation energies corresponding to specific experimental conditions.

In conclusion, we used DFT with hybrid functionals to investigate cation vacancies in monoclinic AlGaO alloys. Our calculations show that the lowest formation energy cation vacancies in AlGaO$_3$ and Al$_2$O$_3$ are split-vacancy configurations, similar to those in Ga$_2$O$_3$. These vacancies have low formation energies for Fermi levels close to the conduction band. In O-poor conditions, Once the alloy reaches an Al content around 16\%, this formation energy will become lower than that of a Si donor, implying that the vacancies will form in large enough concentrations to completely compensate intentional $n$-type doping in these alloys. This effect is more prominent in O-rich growth conditions, as these conditions lead to lower vacancy formation energies compared to O-poor conditions, resulting in vacancies that are always lower in energy than Si near the CBM. O-poor growth conditions can therefore partially mitigate the vacancy compensation. These results explain the experimental observation of electrical compensation in these alloys, as these vacancies are likely to form in larger quantities than Si under most conditions. 

\section*{Supplementary Material}
Details of how the chemical potentials were calculated and a table listing the exact values used are given in the Supplementary Material. 

\section*{Acknowledgments}
This research was supported by the University of Kansas General Research Fund (2151089) and by the National Science Foundation (NSF) through grant DMR-2425549. Computational resources were provided by Bridges-2 at Pittsburgh Supercomputing Center through allocation DMR200079 from the Advanced Cyberinfrastructure Coordination Ecosystem: Services \& Support (ACCESS) program, which is supported by NSF grants \#2138259, \#2138286, \#2138307, \#2137603, and \#2138296, and by the University of Kansas Center for Research Computing (CRC), including the BigJay Cluster resource funded through NSF grant MRI-2117449.

\section*{Data Availability Statement}
The data that support the findings of this study are available from the corresponding author upon reasonable request.

\section*{References}
\bibliography{algao_vac}

\begin{thebibliography}{47}%
\makeatletter
\providecommand \@ifxundefined [1]{%
 \@ifx{#1\undefined}
}%
\providecommand \@ifnum [1]{%
 \ifnum #1\expandafter \@firstoftwo
 \else \expandafter \@secondoftwo
 \fi
}%
\providecommand \@ifx [1]{%
 \ifx #1\expandafter \@firstoftwo
 \else \expandafter \@secondoftwo
 \fi
}%
\providecommand \natexlab [1]{#1}%
\providecommand \enquote  [1]{``#1''}%
\providecommand \bibnamefont  [1]{#1}%
\providecommand \bibfnamefont [1]{#1}%
\providecommand \citenamefont [1]{#1}%
\providecommand \href@noop [0]{\@secondoftwo}%
\providecommand \href [0]{\begingroup \@sanitize@url \@href}%
\providecommand \@href[1]{\@@startlink{#1}\@@href}%
\providecommand \@@href[1]{\endgroup#1\@@endlink}%
\providecommand \@sanitize@url [0]{\catcode `\\12\catcode `\$12\catcode
  `\&12\catcode `\#12\catcode `\^12\catcode `\_12\catcode `\%12\relax}%
\providecommand \@@startlink[1]{}%
\providecommand \@@endlink[0]{}%
\providecommand \url  [0]{\begingroup\@sanitize@url \@url }%
\providecommand \@url [1]{\endgroup\@href {#1}{\urlprefix }}%
\providecommand \urlprefix  [0]{URL }%
\providecommand \Eprint [0]{\href }%
\providecommand \doibase [0]{http://dx.doi.org/}%
\providecommand \selectlanguage [0]{\@gobble}%
\providecommand \bibinfo  [0]{\@secondoftwo}%
\providecommand \bibfield  [0]{\@secondoftwo}%
\providecommand \translation [1]{[#1]}%
\providecommand \BibitemOpen [0]{}%
\providecommand \bibitemStop [0]{}%
\providecommand \bibitemNoStop [0]{.\EOS\space}%
\providecommand \EOS [0]{\spacefactor3000\relax}%
\providecommand \BibitemShut  [1]{\csname bibitem#1\endcsname}%
\let\auto@bib@innerbib\@empty
\bibitem [{\citenamefont {Tippins}(1965)}]{Tippins1965}%
  \BibitemOpen
  \bibfield  {author} {\bibinfo {author} {\bibfnamefont {H.~H.}\ \bibnamefont
  {Tippins}},\ }\href {\doibase 10.1103/PhysRev.140.A316} {\bibfield  {journal}
  {\bibinfo  {journal} {Phys. Rev.}\ }\textbf {\bibinfo {volume} {140}},\
  \bibinfo {pages} {A316} (\bibinfo {year} {1965})}\BibitemShut {NoStop}%
\bibitem [{\citenamefont {Matsumoto}\ \emph {et~al.}(1974)\citenamefont
  {Matsumoto}, \citenamefont {Aoki}, \citenamefont {Kinoshita},\ and\
  \citenamefont {Aono}}]{Matsumoto1974}%
  \BibitemOpen
  \bibfield  {author} {\bibinfo {author} {\bibfnamefont {T.}~\bibnamefont
  {Matsumoto}}, \bibinfo {author} {\bibfnamefont {M.}~\bibnamefont {Aoki}},
  \bibinfo {author} {\bibfnamefont {A.}~\bibnamefont {Kinoshita}}, \ and\
  \bibinfo {author} {\bibfnamefont {T.}~\bibnamefont {Aono}},\ }\href {\doibase
  10.1143/JJAP.13.737} {\bibfield  {journal} {\bibinfo  {journal} {Jpn. J.
  Appl. Phys.}\ }\textbf {\bibinfo {volume} {13}},\ \bibinfo {pages} {737}
  (\bibinfo {year} {1974})}\BibitemShut {NoStop}%
\bibitem [{\citenamefont {Pearton}\ \emph
  {et~al.}(2018{\natexlab{a}})\citenamefont {Pearton}, \citenamefont {Yang},
  \citenamefont {Cary}, \citenamefont {Ren}, \citenamefont {Kim}, \citenamefont
  {Tadjer},\ and\ \citenamefont {Mastro}}]{Pearton2018a}%
  \BibitemOpen
  \bibfield  {author} {\bibinfo {author} {\bibfnamefont {S.~J.}\ \bibnamefont
  {Pearton}}, \bibinfo {author} {\bibfnamefont {J.}~\bibnamefont {Yang}},
  \bibinfo {author} {\bibfnamefont {P.~H.}\ \bibnamefont {Cary}}, \bibinfo
  {author} {\bibfnamefont {F.}~\bibnamefont {Ren}}, \bibinfo {author}
  {\bibfnamefont {J.}~\bibnamefont {Kim}}, \bibinfo {author} {\bibfnamefont
  {M.~J.}\ \bibnamefont {Tadjer}}, \ and\ \bibinfo {author} {\bibfnamefont
  {M.~A.}\ \bibnamefont {Mastro}},\ }\href {\doibase 10.1063/1.5006941}
  {\bibfield  {journal} {\bibinfo  {journal} {Applied Physics Reviews}\
  }\textbf {\bibinfo {volume} {5}},\ \bibinfo {pages} {011301} (\bibinfo {year}
  {2018}{\natexlab{a}})}\BibitemShut {NoStop}%
\bibitem [{\citenamefont {Higashiwaki}\ \emph {et~al.}(2017)\citenamefont
  {Higashiwaki}, \citenamefont {Kuramata}, \citenamefont {Murakami},\ and\
  \citenamefont {Kumagai}}]{Higashiwaki2017}%
  \BibitemOpen
  \bibfield  {author} {\bibinfo {author} {\bibfnamefont {M.}~\bibnamefont
  {Higashiwaki}}, \bibinfo {author} {\bibfnamefont {A.}~\bibnamefont
  {Kuramata}}, \bibinfo {author} {\bibfnamefont {H.}~\bibnamefont {Murakami}},
  \ and\ \bibinfo {author} {\bibfnamefont {Y.}~\bibnamefont {Kumagai}},\ }\href
  {\doibase 10.1088/1361-6463/aa7aff} {\bibfield  {journal} {\bibinfo
  {journal} {J. Phys. D: Appl. Phys.}\ }\textbf {\bibinfo {volume} {50}},\
  \bibinfo {pages} {333002} (\bibinfo {year} {2017})}\BibitemShut {NoStop}%
\bibitem [{\citenamefont {Ji}\ \emph {et~al.}(2022)\citenamefont {Ji},
  \citenamefont {Lu}, \citenamefont {Yan}, \citenamefont {Shan}, \citenamefont
  {Yan}, \citenamefont {Wang}, \citenamefont {Yue}, \citenamefont {Qi},
  \citenamefont {Liu}, \citenamefont {Tang},\ and\ \citenamefont
  {Li}}]{Ji2022}%
  \BibitemOpen
  \bibfield  {author} {\bibinfo {author} {\bibfnamefont {X.}~\bibnamefont
  {Ji}}, \bibinfo {author} {\bibfnamefont {C.}~\bibnamefont {Lu}}, \bibinfo
  {author} {\bibfnamefont {Z.}~\bibnamefont {Yan}}, \bibinfo {author}
  {\bibfnamefont {L.}~\bibnamefont {Shan}}, \bibinfo {author} {\bibfnamefont
  {X.}~\bibnamefont {Yan}}, \bibinfo {author} {\bibfnamefont {J.}~\bibnamefont
  {Wang}}, \bibinfo {author} {\bibfnamefont {J.}~\bibnamefont {Yue}}, \bibinfo
  {author} {\bibfnamefont {X.}~\bibnamefont {Qi}}, \bibinfo {author}
  {\bibfnamefont {Z.}~\bibnamefont {Liu}}, \bibinfo {author} {\bibfnamefont
  {W.}~\bibnamefont {Tang}}, \ and\ \bibinfo {author} {\bibfnamefont
  {P.}~\bibnamefont {Li}},\ }\href {\doibase 10.1088/1361-6463/ac855c}
  {\bibfield  {journal} {\bibinfo  {journal} {J. Phys. D: Appl. Phys.}\
  }\textbf {\bibinfo {volume} {55}},\ \bibinfo {pages} {443002} (\bibinfo
  {year} {2022})}\BibitemShut {NoStop}%
\bibitem [{\citenamefont {Pearton}\ \emph
  {et~al.}(2018{\natexlab{b}})\citenamefont {Pearton}, \citenamefont {Ren},
  \citenamefont {Tadjer},\ and\ \citenamefont {Kim}}]{Pearton2018b}%
  \BibitemOpen
  \bibfield  {author} {\bibinfo {author} {\bibfnamefont {S.~J.}\ \bibnamefont
  {Pearton}}, \bibinfo {author} {\bibfnamefont {F.}~\bibnamefont {Ren}},
  \bibinfo {author} {\bibfnamefont {M.}~\bibnamefont {Tadjer}}, \ and\ \bibinfo
  {author} {\bibfnamefont {J.}~\bibnamefont {Kim}},\ }\href {\doibase
  10.1063/1.5062841} {\bibfield  {journal} {\bibinfo  {journal} {Journal of
  Applied Physics}\ }\textbf {\bibinfo {volume} {124}},\ \bibinfo {pages}
  {220901} (\bibinfo {year} {2018}{\natexlab{b}})}\BibitemShut {NoStop}%
\bibitem [{\citenamefont {Zhou}\ \emph {et~al.}(2019)\citenamefont {Zhou},
  \citenamefont {Zhang}, \citenamefont {Zhang}, \citenamefont {Feng},
  \citenamefont {Zhao}, \citenamefont {Ma},\ and\ \citenamefont
  {Hao}}]{Zhou2019}%
  \BibitemOpen
  \bibfield  {author} {\bibinfo {author} {\bibfnamefont {H.}~\bibnamefont
  {Zhou}}, \bibinfo {author} {\bibfnamefont {J.}~\bibnamefont {Zhang}},
  \bibinfo {author} {\bibfnamefont {C.}~\bibnamefont {Zhang}}, \bibinfo
  {author} {\bibfnamefont {Q.}~\bibnamefont {Feng}}, \bibinfo {author}
  {\bibfnamefont {S.}~\bibnamefont {Zhao}}, \bibinfo {author} {\bibfnamefont
  {P.}~\bibnamefont {Ma}}, \ and\ \bibinfo {author} {\bibfnamefont
  {Y.}~\bibnamefont {Hao}},\ }\href {\doibase 10.1088/1674-4926/40/1/011803}
  {\bibfield  {journal} {\bibinfo  {journal} {J. Semicond.}\ }\textbf {\bibinfo
  {volume} {40}},\ \bibinfo {pages} {011803} (\bibinfo {year} {2019})},\
  \bibinfo {note} {publisher: Chinese Institute of Electronics}\BibitemShut
  {NoStop}%
\bibitem [{\citenamefont {Ueda}\ \emph {et~al.}(1997)\citenamefont {Ueda},
  \citenamefont {Hosono}, \citenamefont {Waseda},\ and\ \citenamefont
  {Kawazoe}}]{Ueda1997}%
  \BibitemOpen
  \bibfield  {author} {\bibinfo {author} {\bibfnamefont {N.}~\bibnamefont
  {Ueda}}, \bibinfo {author} {\bibfnamefont {H.}~\bibnamefont {Hosono}},
  \bibinfo {author} {\bibfnamefont {R.}~\bibnamefont {Waseda}}, \ and\ \bibinfo
  {author} {\bibfnamefont {H.}~\bibnamefont {Kawazoe}},\ }\href {\doibase
  10.1063/1.119233} {\bibfield  {journal} {\bibinfo  {journal} {Applied Physics
  Letters}\ }\textbf {\bibinfo {volume} {70}},\ \bibinfo {pages} {3561}
  (\bibinfo {year} {1997})}\BibitemShut {NoStop}%
\bibitem [{\citenamefont {Orita}\ \emph {et~al.}(2000)\citenamefont {Orita},
  \citenamefont {Ohta}, \citenamefont {Hirano},\ and\ \citenamefont
  {Hosono}}]{Orita2000}%
  \BibitemOpen
  \bibfield  {author} {\bibinfo {author} {\bibfnamefont {M.}~\bibnamefont
  {Orita}}, \bibinfo {author} {\bibfnamefont {H.}~\bibnamefont {Ohta}},
  \bibinfo {author} {\bibfnamefont {M.}~\bibnamefont {Hirano}}, \ and\ \bibinfo
  {author} {\bibfnamefont {H.}~\bibnamefont {Hosono}},\ }\href {\doibase
  10.1063/1.1330559} {\bibfield  {journal} {\bibinfo  {journal} {Applied
  Physics Letters}\ }\textbf {\bibinfo {volume} {77}},\ \bibinfo {pages} {4166}
  (\bibinfo {year} {2000})}\BibitemShut {NoStop}%
\bibitem [{\citenamefont {Guo}\ \emph {et~al.}(2019)\citenamefont {Guo},
  \citenamefont {Guo}, \citenamefont {Chen}, \citenamefont {Wu}, \citenamefont
  {Li},\ and\ \citenamefont {Tang}}]{Guo2019}%
  \BibitemOpen
  \bibfield  {author} {\bibinfo {author} {\bibfnamefont {D.}~\bibnamefont
  {Guo}}, \bibinfo {author} {\bibfnamefont {Q.}~\bibnamefont {Guo}}, \bibinfo
  {author} {\bibfnamefont {Z.}~\bibnamefont {Chen}}, \bibinfo {author}
  {\bibfnamefont {Z.}~\bibnamefont {Wu}}, \bibinfo {author} {\bibfnamefont
  {P.}~\bibnamefont {Li}}, \ and\ \bibinfo {author} {\bibfnamefont
  {W.}~\bibnamefont {Tang}},\ }\href {\doibase 10.1016/j.mtphys.2019.100157}
  {\bibfield  {journal} {\bibinfo  {journal} {Mater. Today Phys.}\ }\textbf
  {\bibinfo {volume} {11}},\ \bibinfo {pages} {100157} (\bibinfo {year}
  {2019})}\BibitemShut {NoStop}%
\bibitem [{\citenamefont {Wang}\ \emph {et~al.}(2021)\citenamefont {Wang},
  \citenamefont {Zhang}, \citenamefont {Xu}, \citenamefont {Zhang},
  \citenamefont {Feng}, \citenamefont {Zhang}, \citenamefont {Ning},
  \citenamefont {Zhao}, \citenamefont {Zhou},\ and\ \citenamefont
  {Hao}}]{Wang2021}%
  \BibitemOpen
  \bibfield  {author} {\bibinfo {author} {\bibfnamefont {C.}~\bibnamefont
  {Wang}}, \bibinfo {author} {\bibfnamefont {J.}~\bibnamefont {Zhang}},
  \bibinfo {author} {\bibfnamefont {S.}~\bibnamefont {Xu}}, \bibinfo {author}
  {\bibfnamefont {C.}~\bibnamefont {Zhang}}, \bibinfo {author} {\bibfnamefont
  {Q.}~\bibnamefont {Feng}}, \bibinfo {author} {\bibfnamefont {Y.}~\bibnamefont
  {Zhang}}, \bibinfo {author} {\bibfnamefont {J.}~\bibnamefont {Ning}},
  \bibinfo {author} {\bibfnamefont {S.}~\bibnamefont {Zhao}}, \bibinfo {author}
  {\bibfnamefont {H.}~\bibnamefont {Zhou}}, \ and\ \bibinfo {author}
  {\bibfnamefont {Y.}~\bibnamefont {Hao}},\ }\href {\doibase
  10.1088/1361-6463/abe158} {\bibfield  {journal} {\bibinfo  {journal} {J.
  Phys. D: Appl. Phys.}\ }\textbf {\bibinfo {volume} {54}},\ \bibinfo {pages}
  {243001} (\bibinfo {year} {2021})}\BibitemShut {NoStop}%
\bibitem [{\citenamefont {Sun}\ \emph {et~al.}(2024)\citenamefont {Sun},
  \citenamefont {Wang}, \citenamefont {Alghamdi}, \citenamefont {Zhou},
  \citenamefont {Hao},\ and\ \citenamefont {Zhang}}]{Sun2024}%
  \BibitemOpen
  \bibfield  {author} {\bibinfo {author} {\bibfnamefont {S.}~\bibnamefont
  {Sun}}, \bibinfo {author} {\bibfnamefont {C.}~\bibnamefont {Wang}}, \bibinfo
  {author} {\bibfnamefont {S.}~\bibnamefont {Alghamdi}}, \bibinfo {author}
  {\bibfnamefont {H.}~\bibnamefont {Zhou}}, \bibinfo {author} {\bibfnamefont
  {Y.}~\bibnamefont {Hao}}, \ and\ \bibinfo {author} {\bibfnamefont
  {J.}~\bibnamefont {Zhang}},\ }\href {\doibase 10.1002/aelm.202300844}
  {\bibfield  {journal} {\bibinfo  {journal} {Advanced Electronic Materials}\
  }\textbf {\bibinfo {volume} {11}},\ \bibinfo {pages} {2300844} (\bibinfo
  {year} {2024})}\BibitemShut {NoStop}%
\bibitem [{\citenamefont {Ahmadi}\ \emph {et~al.}(2017)\citenamefont {Ahmadi},
  \citenamefont {Koksaldi}, \citenamefont {Zheng}, \citenamefont {Mates},
  \citenamefont {Oshima}, \citenamefont {Mishra},\ and\ \citenamefont
  {Speck}}]{Ahmadi2017a}%
  \BibitemOpen
  \bibfield  {author} {\bibinfo {author} {\bibfnamefont {E.}~\bibnamefont
  {Ahmadi}}, \bibinfo {author} {\bibfnamefont {O.~S.}\ \bibnamefont
  {Koksaldi}}, \bibinfo {author} {\bibfnamefont {X.}~\bibnamefont {Zheng}},
  \bibinfo {author} {\bibfnamefont {T.}~\bibnamefont {Mates}}, \bibinfo
  {author} {\bibfnamefont {Y.}~\bibnamefont {Oshima}}, \bibinfo {author}
  {\bibfnamefont {U.~K.}\ \bibnamefont {Mishra}}, \ and\ \bibinfo {author}
  {\bibfnamefont {J.~S.}\ \bibnamefont {Speck}},\ }\href {\doibase
  10.7567/APEX.10.071101} {\bibfield  {journal} {\bibinfo  {journal} {Appl.
  Phys. Express}\ }\textbf {\bibinfo {volume} {10}},\ \bibinfo {pages} {071101}
  (\bibinfo {year} {2017})}\BibitemShut {NoStop}%
\bibitem [{\citenamefont {Krishnamoorthy}\ \emph {et~al.}(2017)\citenamefont
  {Krishnamoorthy}, \citenamefont {Xia}, \citenamefont {Joishi}, \citenamefont
  {Zhang}, \citenamefont {McGlone}, \citenamefont {Johnson}, \citenamefont
  {Brenner}, \citenamefont {Arehart}, \citenamefont {Hwang}, \citenamefont
  {Lodha},\ and\ \citenamefont {Rajan}}]{Krishnamoorthy2017}%
  \BibitemOpen
  \bibfield  {author} {\bibinfo {author} {\bibfnamefont {S.}~\bibnamefont
  {Krishnamoorthy}}, \bibinfo {author} {\bibfnamefont {Z.}~\bibnamefont {Xia}},
  \bibinfo {author} {\bibfnamefont {C.}~\bibnamefont {Joishi}}, \bibinfo
  {author} {\bibfnamefont {Y.}~\bibnamefont {Zhang}}, \bibinfo {author}
  {\bibfnamefont {J.}~\bibnamefont {McGlone}}, \bibinfo {author} {\bibfnamefont
  {J.}~\bibnamefont {Johnson}}, \bibinfo {author} {\bibfnamefont
  {M.}~\bibnamefont {Brenner}}, \bibinfo {author} {\bibfnamefont {A.~R.}\
  \bibnamefont {Arehart}}, \bibinfo {author} {\bibfnamefont {J.}~\bibnamefont
  {Hwang}}, \bibinfo {author} {\bibfnamefont {S.}~\bibnamefont {Lodha}}, \ and\
  \bibinfo {author} {\bibfnamefont {S.}~\bibnamefont {Rajan}},\ }\href
  {\doibase 10.1063/1.4993569} {\bibfield  {journal} {\bibinfo  {journal}
  {Appl. Phys. Lett.}\ }\textbf {\bibinfo {volume} {111}},\ \bibinfo {pages}
  {023502} (\bibinfo {year} {2017})}\BibitemShut {NoStop}%
\bibitem [{\citenamefont {Joishi}\ \emph {et~al.}(2019)\citenamefont {Joishi},
  \citenamefont {Zhang}, \citenamefont {Xia}, \citenamefont {Sun},
  \citenamefont {Arehart}, \citenamefont {Ringel}, \citenamefont {Lodha},\ and\
  \citenamefont {Rajan}}]{Joishi2019}%
  \BibitemOpen
  \bibfield  {author} {\bibinfo {author} {\bibfnamefont {C.}~\bibnamefont
  {Joishi}}, \bibinfo {author} {\bibfnamefont {Y.}~\bibnamefont {Zhang}},
  \bibinfo {author} {\bibfnamefont {Z.}~\bibnamefont {Xia}}, \bibinfo {author}
  {\bibfnamefont {W.}~\bibnamefont {Sun}}, \bibinfo {author} {\bibfnamefont
  {A.~R.}\ \bibnamefont {Arehart}}, \bibinfo {author} {\bibfnamefont
  {S.}~\bibnamefont {Ringel}}, \bibinfo {author} {\bibfnamefont
  {S.}~\bibnamefont {Lodha}}, \ and\ \bibinfo {author} {\bibfnamefont
  {S.}~\bibnamefont {Rajan}},\ }\href {\doibase 10.1109/LED.2019.2921116}
  {\bibfield  {journal} {\bibinfo  {journal} {IEEE Electron Device Lett.}\
  }\textbf {\bibinfo {volume} {40}},\ \bibinfo {pages} {1241} (\bibinfo {year}
  {2019})}\BibitemShut {NoStop}%
\bibitem [{\citenamefont {Okumura}\ \emph {et~al.}(2019)\citenamefont
  {Okumura}, \citenamefont {Kato}, \citenamefont {Oshima},\ and\ \citenamefont
  {Palacios}}]{Okumura2019}%
  \BibitemOpen
  \bibfield  {author} {\bibinfo {author} {\bibfnamefont {H.}~\bibnamefont
  {Okumura}}, \bibinfo {author} {\bibfnamefont {Y.}~\bibnamefont {Kato}},
  \bibinfo {author} {\bibfnamefont {T.}~\bibnamefont {Oshima}}, \ and\ \bibinfo
  {author} {\bibfnamefont {T.}~\bibnamefont {Palacios}},\ }\href {\doibase
  10.7567/1347-4065/ab002b} {\bibfield  {journal} {\bibinfo  {journal} {Jpn. J.
  Appl. Phys.}\ }\textbf {\bibinfo {volume} {58}},\ \bibinfo {pages} {SBBD12}
  (\bibinfo {year} {2019})}\BibitemShut {NoStop}%
\bibitem [{\citenamefont {Chatterjee}\ \emph {et~al.}(2020)\citenamefont
  {Chatterjee}, \citenamefont {Song}, \citenamefont {Lundh}, \citenamefont
  {Zhang}, \citenamefont {Xia}, \citenamefont {Islam}, \citenamefont {Leach},
  \citenamefont {McGray}, \citenamefont {Ranga}, \citenamefont
  {Krishnamoorthy}, \citenamefont {Haque}, \citenamefont {Rajan},\ and\
  \citenamefont {Choi}}]{Chatterjee2020}%
  \BibitemOpen
  \bibfield  {author} {\bibinfo {author} {\bibfnamefont {B.}~\bibnamefont
  {Chatterjee}}, \bibinfo {author} {\bibfnamefont {Y.}~\bibnamefont {Song}},
  \bibinfo {author} {\bibfnamefont {J.~S.}\ \bibnamefont {Lundh}}, \bibinfo
  {author} {\bibfnamefont {Y.}~\bibnamefont {Zhang}}, \bibinfo {author}
  {\bibfnamefont {Z.}~\bibnamefont {Xia}}, \bibinfo {author} {\bibfnamefont
  {Z.}~\bibnamefont {Islam}}, \bibinfo {author} {\bibfnamefont
  {J.}~\bibnamefont {Leach}}, \bibinfo {author} {\bibfnamefont
  {C.}~\bibnamefont {McGray}}, \bibinfo {author} {\bibfnamefont
  {P.}~\bibnamefont {Ranga}}, \bibinfo {author} {\bibfnamefont
  {S.}~\bibnamefont {Krishnamoorthy}}, \bibinfo {author} {\bibfnamefont
  {A.}~\bibnamefont {Haque}}, \bibinfo {author} {\bibfnamefont
  {S.}~\bibnamefont {Rajan}}, \ and\ \bibinfo {author} {\bibfnamefont
  {S.}~\bibnamefont {Choi}},\ }\href {\doibase 10.1063/5.0021275} {\bibfield
  {journal} {\bibinfo  {journal} {Appl. Phys. Lett.}\ }\textbf {\bibinfo
  {volume} {117}},\ \bibinfo {pages} {153501} (\bibinfo {year}
  {2020})}\BibitemShut {NoStop}%
\bibitem [{\citenamefont {Vaidya}, \citenamefont {Saha},\ and\ \citenamefont
  {Singisetti}(2021)}]{Vaidya2021}%
  \BibitemOpen
  \bibfield  {author} {\bibinfo {author} {\bibfnamefont {A.}~\bibnamefont
  {Vaidya}}, \bibinfo {author} {\bibfnamefont {C.~N.}\ \bibnamefont {Saha}}, \
  and\ \bibinfo {author} {\bibfnamefont {U.}~\bibnamefont {Singisetti}},\
  }\href {\doibase 10.1109/LED.2021.3104256} {\bibfield  {journal} {\bibinfo
  {journal} {IEEE Electron Device Lett.}\ }\textbf {\bibinfo {volume} {42}},\
  \bibinfo {pages} {1444} (\bibinfo {year} {2021})}\BibitemShut {NoStop}%
\bibitem [{\citenamefont {Tadjer}\ \emph {et~al.}(2021)\citenamefont {Tadjer},
  \citenamefont {Sasaki}, \citenamefont {Wakimoto}, \citenamefont {Anderson},
  \citenamefont {Mastro}, \citenamefont {Gallagher}, \citenamefont {Jacobs},
  \citenamefont {Mock}, \citenamefont {Koehler}, \citenamefont {Ebrish},
  \citenamefont {Hobart},\ and\ \citenamefont {Kuramata}}]{Tadjer2021}%
  \BibitemOpen
  \bibfield  {author} {\bibinfo {author} {\bibfnamefont {M.~J.}\ \bibnamefont
  {Tadjer}}, \bibinfo {author} {\bibfnamefont {K.}~\bibnamefont {Sasaki}},
  \bibinfo {author} {\bibfnamefont {D.}~\bibnamefont {Wakimoto}}, \bibinfo
  {author} {\bibfnamefont {T.~J.}\ \bibnamefont {Anderson}}, \bibinfo {author}
  {\bibfnamefont {M.~A.}\ \bibnamefont {Mastro}}, \bibinfo {author}
  {\bibfnamefont {J.~C.}\ \bibnamefont {Gallagher}}, \bibinfo {author}
  {\bibfnamefont {A.~G.}\ \bibnamefont {Jacobs}}, \bibinfo {author}
  {\bibfnamefont {A.~L.}\ \bibnamefont {Mock}}, \bibinfo {author}
  {\bibfnamefont {A.~D.}\ \bibnamefont {Koehler}}, \bibinfo {author}
  {\bibfnamefont {M.}~\bibnamefont {Ebrish}}, \bibinfo {author} {\bibfnamefont
  {K.~D.}\ \bibnamefont {Hobart}}, \ and\ \bibinfo {author} {\bibfnamefont
  {A.}~\bibnamefont {Kuramata}},\ }\href {\doibase 10.1116/6.0000932}
  {\bibfield  {journal} {\bibinfo  {journal} {J. Vac. Sci. Technol. A}\
  }\textbf {\bibinfo {volume} {39}},\ \bibinfo {pages} {033402} (\bibinfo
  {year} {2021})}\BibitemShut {NoStop}%
\bibitem [{\citenamefont {Peelaers}\ \emph {et~al.}(2018)\citenamefont
  {Peelaers}, \citenamefont {Varley}, \citenamefont {Speck},\ and\
  \citenamefont {Van~de Walle}}]{Peelaers2018}%
  \BibitemOpen
  \bibfield  {author} {\bibinfo {author} {\bibfnamefont {H.}~\bibnamefont
  {Peelaers}}, \bibinfo {author} {\bibfnamefont {J.~B.}\ \bibnamefont
  {Varley}}, \bibinfo {author} {\bibfnamefont {J.~S.}\ \bibnamefont {Speck}}, \
  and\ \bibinfo {author} {\bibfnamefont {C.~G.}\ \bibnamefont {Van~de Walle}},\
  }\href {\doibase 10.1063/1.5036991} {\bibfield  {journal} {\bibinfo
  {journal} {Appl. Phys. Lett.}\ }\textbf {\bibinfo {volume} {112}},\ \bibinfo
  {pages} {242101} (\bibinfo {year} {2018})}\BibitemShut {NoStop}%
\bibitem [{\citenamefont {Zhang}\ \emph {et~al.}(2018)\citenamefont {Zhang},
  \citenamefont {Neal}, \citenamefont {Xia}, \citenamefont {Joishi},
  \citenamefont {Johnson}, \citenamefont {Zheng}, \citenamefont {Bajaj},
  \citenamefont {Brenner}, \citenamefont {Dorsey}, \citenamefont {Chabak},
  \citenamefont {Jessen}, \citenamefont {Hwang}, \citenamefont {Mou},
  \citenamefont {Heremans},\ and\ \citenamefont {Rajan}}]{Zhang2018}%
  \BibitemOpen
  \bibfield  {author} {\bibinfo {author} {\bibfnamefont {Y.}~\bibnamefont
  {Zhang}}, \bibinfo {author} {\bibfnamefont {A.}~\bibnamefont {Neal}},
  \bibinfo {author} {\bibfnamefont {Z.}~\bibnamefont {Xia}}, \bibinfo {author}
  {\bibfnamefont {C.}~\bibnamefont {Joishi}}, \bibinfo {author} {\bibfnamefont
  {J.~M.}\ \bibnamefont {Johnson}}, \bibinfo {author} {\bibfnamefont
  {Y.}~\bibnamefont {Zheng}}, \bibinfo {author} {\bibfnamefont
  {S.}~\bibnamefont {Bajaj}}, \bibinfo {author} {\bibfnamefont
  {M.}~\bibnamefont {Brenner}}, \bibinfo {author} {\bibfnamefont
  {D.}~\bibnamefont {Dorsey}}, \bibinfo {author} {\bibfnamefont
  {K.}~\bibnamefont {Chabak}}, \bibinfo {author} {\bibfnamefont
  {G.}~\bibnamefont {Jessen}}, \bibinfo {author} {\bibfnamefont
  {J.}~\bibnamefont {Hwang}}, \bibinfo {author} {\bibfnamefont
  {S.}~\bibnamefont {Mou}}, \bibinfo {author} {\bibfnamefont {J.~P.}\
  \bibnamefont {Heremans}}, \ and\ \bibinfo {author} {\bibfnamefont
  {S.}~\bibnamefont {Rajan}},\ }\href {\doibase 10.1063/1.5025704} {\bibfield
  {journal} {\bibinfo  {journal} {Applied Physics Letters}\ }\textbf {\bibinfo
  {volume} {112}},\ \bibinfo {pages} {173502} (\bibinfo {year}
  {2018})}\BibitemShut {NoStop}%
\bibitem [{\citenamefont {Ranga}\ \emph {et~al.}(2021)\citenamefont {Ranga},
  \citenamefont {Bhattacharyya}, \citenamefont {Chmielewski}, \citenamefont
  {Roy}, \citenamefont {Sun}, \citenamefont {Scarpulla}, \citenamefont {Alem},\
  and\ \citenamefont {Krishnamoorthy}}]{Ranga2021}%
  \BibitemOpen
  \bibfield  {author} {\bibinfo {author} {\bibfnamefont {P.}~\bibnamefont
  {Ranga}}, \bibinfo {author} {\bibfnamefont {A.}~\bibnamefont
  {Bhattacharyya}}, \bibinfo {author} {\bibfnamefont {A.}~\bibnamefont
  {Chmielewski}}, \bibinfo {author} {\bibfnamefont {S.}~\bibnamefont {Roy}},
  \bibinfo {author} {\bibfnamefont {R.}~\bibnamefont {Sun}}, \bibinfo {author}
  {\bibfnamefont {M.~A.}\ \bibnamefont {Scarpulla}}, \bibinfo {author}
  {\bibfnamefont {N.}~\bibnamefont {Alem}}, \ and\ \bibinfo {author}
  {\bibfnamefont {S.}~\bibnamefont {Krishnamoorthy}},\ }\href {\doibase
  10.35848/1882-0786/abd675} {\bibfield  {journal} {\bibinfo  {journal} {Appl.
  Phys. Express}\ }\textbf {\bibinfo {volume} {14}},\ \bibinfo {pages} {025501}
  (\bibinfo {year} {2021})}\BibitemShut {NoStop}%
\bibitem [{\citenamefont {Galazka}\ \emph {et~al.}(2023)\citenamefont
  {Galazka}, \citenamefont {Fiedler}, \citenamefont {Popp}, \citenamefont
  {Ganschow}, \citenamefont {Kwasniewski}, \citenamefont {Seyidov},
  \citenamefont {Pietsch}, \citenamefont {Dittmar}, \citenamefont {Anooz},
  \citenamefont {Irmscher}, \citenamefont {Suendermann}, \citenamefont {Klimm},
  \citenamefont {Chou}, \citenamefont {Rehm}, \citenamefont {Schroeder},\ and\
  \citenamefont {Bickermann}}]{Galazka2023}%
  \BibitemOpen
  \bibfield  {author} {\bibinfo {author} {\bibfnamefont {Z.}~\bibnamefont
  {Galazka}}, \bibinfo {author} {\bibfnamefont {A.}~\bibnamefont {Fiedler}},
  \bibinfo {author} {\bibfnamefont {A.}~\bibnamefont {Popp}}, \bibinfo {author}
  {\bibfnamefont {S.}~\bibnamefont {Ganschow}}, \bibinfo {author}
  {\bibfnamefont {A.}~\bibnamefont {Kwasniewski}}, \bibinfo {author}
  {\bibfnamefont {P.}~\bibnamefont {Seyidov}}, \bibinfo {author} {\bibfnamefont
  {M.}~\bibnamefont {Pietsch}}, \bibinfo {author} {\bibfnamefont
  {A.}~\bibnamefont {Dittmar}}, \bibinfo {author} {\bibfnamefont {S.~B.}\
  \bibnamefont {Anooz}}, \bibinfo {author} {\bibfnamefont {K.}~\bibnamefont
  {Irmscher}}, \bibinfo {author} {\bibfnamefont {M.}~\bibnamefont
  {Suendermann}}, \bibinfo {author} {\bibfnamefont {D.}~\bibnamefont {Klimm}},
  \bibinfo {author} {\bibfnamefont {T.-S.}\ \bibnamefont {Chou}}, \bibinfo
  {author} {\bibfnamefont {J.}~\bibnamefont {Rehm}}, \bibinfo {author}
  {\bibfnamefont {T.}~\bibnamefont {Schroeder}}, \ and\ \bibinfo {author}
  {\bibfnamefont {M.}~\bibnamefont {Bickermann}},\ }\href {\doibase
  10.1063/5.0131285} {\bibfield  {journal} {\bibinfo  {journal} {Journal of
  Applied Physics}\ }\textbf {\bibinfo {volume} {133}},\ \bibinfo {pages}
  {035702} (\bibinfo {year} {2023})}\BibitemShut {NoStop}%
\bibitem [{\citenamefont {Bhuiyan}\ \emph {et~al.}(2023)\citenamefont
  {Bhuiyan}, \citenamefont {Meng}, \citenamefont {Huang}, \citenamefont {Chae},
  \citenamefont {Hwang},\ and\ \citenamefont {Zhao}}]{Bhuiyan2023}%
  \BibitemOpen
  \bibfield  {author} {\bibinfo {author} {\bibfnamefont {A.~F. M. A.~U.}\
  \bibnamefont {Bhuiyan}}, \bibinfo {author} {\bibfnamefont {L.}~\bibnamefont
  {Meng}}, \bibinfo {author} {\bibfnamefont {H.-L.}\ \bibnamefont {Huang}},
  \bibinfo {author} {\bibfnamefont {C.}~\bibnamefont {Chae}}, \bibinfo {author}
  {\bibfnamefont {J.}~\bibnamefont {Hwang}}, \ and\ \bibinfo {author}
  {\bibfnamefont {H.}~\bibnamefont {Zhao}},\ }\href {\doibase
  10.1002/pssr.202300224} {\bibfield  {journal} {\bibinfo  {journal} {physica
  status solidi (RRL) – Rapid Research Letters}\ }\textbf {\bibinfo {volume}
  {17}},\ \bibinfo {pages} {2300224} (\bibinfo {year} {2023})}\BibitemShut
  {NoStop}%
\bibitem [{\citenamefont {Varley}\ \emph {et~al.}(2020)\citenamefont {Varley},
  \citenamefont {Perron}, \citenamefont {Lordi}, \citenamefont
  {Wickramaratne},\ and\ \citenamefont {Lyons}}]{Varley2020}%
  \BibitemOpen
  \bibfield  {author} {\bibinfo {author} {\bibfnamefont {J.~B.}\ \bibnamefont
  {Varley}}, \bibinfo {author} {\bibfnamefont {A.}~\bibnamefont {Perron}},
  \bibinfo {author} {\bibfnamefont {V.}~\bibnamefont {Lordi}}, \bibinfo
  {author} {\bibfnamefont {D.}~\bibnamefont {Wickramaratne}}, \ and\ \bibinfo
  {author} {\bibfnamefont {J.~L.}\ \bibnamefont {Lyons}},\ }\href {\doibase
  10.1063/5.0006224} {\bibfield  {journal} {\bibinfo  {journal} {Appl. Phys.
  Lett.}\ }\textbf {\bibinfo {volume} {116}},\ \bibinfo {pages} {172104}
  (\bibinfo {year} {2020})}\BibitemShut {NoStop}%
\bibitem [{\citenamefont {Mu}\ \emph {et~al.}(2022)\citenamefont {Mu},
  \citenamefont {Wang}, \citenamefont {Varley}, \citenamefont {Lyons},
  \citenamefont {Wickramaratne},\ and\ \citenamefont {Van De~Walle}}]{Mu2022}%
  \BibitemOpen
  \bibfield  {author} {\bibinfo {author} {\bibfnamefont {S.}~\bibnamefont
  {Mu}}, \bibinfo {author} {\bibfnamefont {M.}~\bibnamefont {Wang}}, \bibinfo
  {author} {\bibfnamefont {J.~B.}\ \bibnamefont {Varley}}, \bibinfo {author}
  {\bibfnamefont {J.~L.}\ \bibnamefont {Lyons}}, \bibinfo {author}
  {\bibfnamefont {D.}~\bibnamefont {Wickramaratne}}, \ and\ \bibinfo {author}
  {\bibfnamefont {C.~G.}\ \bibnamefont {Van De~Walle}},\ }\href {\doibase
  10.1103/PhysRevB.105.155201} {\bibfield  {journal} {\bibinfo  {journal}
  {Phys. Rev. B}\ }\textbf {\bibinfo {volume} {105}},\ \bibinfo {pages}
  {155201} (\bibinfo {year} {2022})}\BibitemShut {NoStop}%
\bibitem [{\citenamefont {Ranga}\ \emph {et~al.}(2020)\citenamefont {Ranga},
  \citenamefont {Bhattacharyya}, \citenamefont {Rishinaramangalam},
  \citenamefont {Ooi}, \citenamefont {Scarpulla}, \citenamefont {Feezell},\
  and\ \citenamefont {Krishnamoorthy}}]{Ranga2020a}%
  \BibitemOpen
  \bibfield  {author} {\bibinfo {author} {\bibfnamefont {P.}~\bibnamefont
  {Ranga}}, \bibinfo {author} {\bibfnamefont {A.}~\bibnamefont
  {Bhattacharyya}}, \bibinfo {author} {\bibfnamefont {A.}~\bibnamefont
  {Rishinaramangalam}}, \bibinfo {author} {\bibfnamefont {Y.~K.}\ \bibnamefont
  {Ooi}}, \bibinfo {author} {\bibfnamefont {M.~A.}\ \bibnamefont {Scarpulla}},
  \bibinfo {author} {\bibfnamefont {D.}~\bibnamefont {Feezell}}, \ and\
  \bibinfo {author} {\bibfnamefont {S.}~\bibnamefont {Krishnamoorthy}},\ }\href
  {\doibase 10.35848/1882-0786/ab7712} {\bibfield  {journal} {\bibinfo
  {journal} {Appl. Phys. Express}\ }\textbf {\bibinfo {volume} {13}},\ \bibinfo
  {pages} {045501} (\bibinfo {year} {2020})}\BibitemShut {NoStop}%
\bibitem [{\citenamefont {Weiser}\ \emph {et~al.}(2018)\citenamefont {Weiser},
  \citenamefont {Stavola}, \citenamefont {Fowler}, \citenamefont {Qin},\ and\
  \citenamefont {Pearton}}]{Weiser2018}%
  \BibitemOpen
  \bibfield  {author} {\bibinfo {author} {\bibfnamefont {P.}~\bibnamefont
  {Weiser}}, \bibinfo {author} {\bibfnamefont {M.}~\bibnamefont {Stavola}},
  \bibinfo {author} {\bibfnamefont {W.~B.}\ \bibnamefont {Fowler}}, \bibinfo
  {author} {\bibfnamefont {Y.}~\bibnamefont {Qin}}, \ and\ \bibinfo {author}
  {\bibfnamefont {S.}~\bibnamefont {Pearton}},\ }\href {\doibase
  10.1063/1.5029921} {\bibfield  {journal} {\bibinfo  {journal} {Applied
  Physics Letters}\ }\textbf {\bibinfo {volume} {112}},\ \bibinfo {pages}
  {232104} (\bibinfo {year} {2018})}\BibitemShut {NoStop}%
\bibitem [{\citenamefont {Son}\ \emph {et~al.}(2020)\citenamefont {Son},
  \citenamefont {Ho}, \citenamefont {Goto}, \citenamefont {Abe}, \citenamefont
  {Ohshima}, \citenamefont {Monemar}, \citenamefont {Kumagai}, \citenamefont
  {Frauenheim},\ and\ \citenamefont {Deák}}]{Son2020}%
  \BibitemOpen
  \bibfield  {author} {\bibinfo {author} {\bibfnamefont {N.~T.}\ \bibnamefont
  {Son}}, \bibinfo {author} {\bibfnamefont {Q.~D.}\ \bibnamefont {Ho}},
  \bibinfo {author} {\bibfnamefont {K.}~\bibnamefont {Goto}}, \bibinfo {author}
  {\bibfnamefont {H.}~\bibnamefont {Abe}}, \bibinfo {author} {\bibfnamefont
  {T.}~\bibnamefont {Ohshima}}, \bibinfo {author} {\bibfnamefont
  {B.}~\bibnamefont {Monemar}}, \bibinfo {author} {\bibfnamefont
  {Y.}~\bibnamefont {Kumagai}}, \bibinfo {author} {\bibfnamefont
  {T.}~\bibnamefont {Frauenheim}}, \ and\ \bibinfo {author} {\bibfnamefont
  {P.}~\bibnamefont {Deák}},\ }\href {\doibase 10.1063/5.0012579} {\bibfield
  {journal} {\bibinfo  {journal} {Applied Physics Letters}\ }\textbf {\bibinfo
  {volume} {117}},\ \bibinfo {pages} {032101} (\bibinfo {year}
  {2020})}\BibitemShut {NoStop}%
\bibitem [{\citenamefont {Karjalainen}\ \emph {et~al.}(2020)\citenamefont
  {Karjalainen}, \citenamefont {Prozheeva}, \citenamefont {Simula},
  \citenamefont {Makkonen}, \citenamefont {Callewaert}, \citenamefont
  {Varley},\ and\ \citenamefont {Tuomisto}}]{Karjalainen2020}%
  \BibitemOpen
  \bibfield  {author} {\bibinfo {author} {\bibfnamefont {A.}~\bibnamefont
  {Karjalainen}}, \bibinfo {author} {\bibfnamefont {V.}~\bibnamefont
  {Prozheeva}}, \bibinfo {author} {\bibfnamefont {K.}~\bibnamefont {Simula}},
  \bibinfo {author} {\bibfnamefont {I.}~\bibnamefont {Makkonen}}, \bibinfo
  {author} {\bibfnamefont {V.}~\bibnamefont {Callewaert}}, \bibinfo {author}
  {\bibfnamefont {J.~B.}\ \bibnamefont {Varley}}, \ and\ \bibinfo {author}
  {\bibfnamefont {F.}~\bibnamefont {Tuomisto}},\ }\href {\doibase
  10.1103/PhysRevB.102.195207} {\bibfield  {journal} {\bibinfo  {journal}
  {Phys. Rev. B}\ }\textbf {\bibinfo {volume} {102}},\ \bibinfo {pages}
  {195207} (\bibinfo {year} {2020})}\BibitemShut {NoStop}%
\bibitem [{\citenamefont {Johnson}\ \emph {et~al.}(2019)\citenamefont
  {Johnson}, \citenamefont {Chen}, \citenamefont {Varley}, \citenamefont
  {Jackson}, \citenamefont {Farzana}, \citenamefont {Zhang}, \citenamefont
  {Arehart}, \citenamefont {Huang}, \citenamefont {Genc}, \citenamefont
  {Ringel}, \citenamefont {Van De~Walle}, \citenamefont {Muller},\ and\
  \citenamefont {Hwang}}]{Johnson2019}%
  \BibitemOpen
  \bibfield  {author} {\bibinfo {author} {\bibfnamefont {J.~M.}\ \bibnamefont
  {Johnson}}, \bibinfo {author} {\bibfnamefont {Z.}~\bibnamefont {Chen}},
  \bibinfo {author} {\bibfnamefont {J.~B.}\ \bibnamefont {Varley}}, \bibinfo
  {author} {\bibfnamefont {C.~M.}\ \bibnamefont {Jackson}}, \bibinfo {author}
  {\bibfnamefont {E.}~\bibnamefont {Farzana}}, \bibinfo {author} {\bibfnamefont
  {Z.}~\bibnamefont {Zhang}}, \bibinfo {author} {\bibfnamefont {A.~R.}\
  \bibnamefont {Arehart}}, \bibinfo {author} {\bibfnamefont {H.-L.}\
  \bibnamefont {Huang}}, \bibinfo {author} {\bibfnamefont {A.}~\bibnamefont
  {Genc}}, \bibinfo {author} {\bibfnamefont {S.~A.}\ \bibnamefont {Ringel}},
  \bibinfo {author} {\bibfnamefont {C.~G.}\ \bibnamefont {Van De~Walle}},
  \bibinfo {author} {\bibfnamefont {D.~A.}\ \bibnamefont {Muller}}, \ and\
  \bibinfo {author} {\bibfnamefont {J.}~\bibnamefont {Hwang}},\ }\href
  {\doibase 10.1103/PhysRevX.9.041027} {\bibfield  {journal} {\bibinfo
  {journal} {Phys. Rev. X}\ }\textbf {\bibinfo {volume} {9}},\ \bibinfo {pages}
  {041027} (\bibinfo {year} {2019})}\BibitemShut {NoStop}%
\bibitem [{\citenamefont {Tuomisto}(2024)}]{Tuomisto2024}%
  \BibitemOpen
  \bibfield  {author} {\bibinfo {author} {\bibfnamefont {F.}~\bibnamefont
  {Tuomisto}},\ }\href {\doibase 10.1557/s43578-024-01407-4} {\bibfield
  {journal} {\bibinfo  {journal} {Journal of Materials Research}\ }\textbf
  {\bibinfo {volume} {39}},\ \bibinfo {pages} {2369} (\bibinfo {year}
  {2024})}\BibitemShut {NoStop}%
\bibitem [{\citenamefont {Varley}\ \emph {et~al.}(2011)\citenamefont {Varley},
  \citenamefont {Peelaers}, \citenamefont {Janotti},\ and\ \citenamefont {{Van
  de Walle}}}]{Varley2011}%
  \BibitemOpen
  \bibfield  {author} {\bibinfo {author} {\bibfnamefont {J.~B.}\ \bibnamefont
  {Varley}}, \bibinfo {author} {\bibfnamefont {H.}~\bibnamefont {Peelaers}},
  \bibinfo {author} {\bibfnamefont {A.}~\bibnamefont {Janotti}}, \ and\
  \bibinfo {author} {\bibfnamefont {C.~G.}\ \bibnamefont {{Van de Walle}}},\
  }\href {\doibase 10.1088/0953-8984/23/33/334212} {\bibfield  {journal}
  {\bibinfo  {journal} {J. Phys. Condens. Matter}\ }\textbf {\bibinfo {volume}
  {23}},\ \bibinfo {pages} {334212} (\bibinfo {year} {2011})}\BibitemShut
  {NoStop}%
\bibitem [{\citenamefont {Kyrtsos}, \citenamefont {Matsubara},\ and\
  \citenamefont {Bellotti}(2017)}]{Kyrtsos2017}%
  \BibitemOpen
  \bibfield  {author} {\bibinfo {author} {\bibfnamefont {A.}~\bibnamefont
  {Kyrtsos}}, \bibinfo {author} {\bibfnamefont {M.}~\bibnamefont {Matsubara}},
  \ and\ \bibinfo {author} {\bibfnamefont {E.}~\bibnamefont {Bellotti}},\
  }\href {\doibase 10.1103/PhysRevB.95.245202} {\bibfield  {journal} {\bibinfo
  {journal} {Phys. Rev. B}\ }\textbf {\bibinfo {volume} {95}},\ \bibinfo
  {pages} {245202} (\bibinfo {year} {2017})}\BibitemShut {NoStop}%
\bibitem [{\citenamefont {Zimmermann}\ \emph {et~al.}(2020)\citenamefont
  {Zimmermann}, \citenamefont {Rønning}, \citenamefont {Kalmann~Frodason},
  \citenamefont {Bobal}, \citenamefont {Vines},\ and\ \citenamefont
  {Varley}}]{Zimmerman2020}%
  \BibitemOpen
  \bibfield  {author} {\bibinfo {author} {\bibfnamefont {C.}~\bibnamefont
  {Zimmermann}}, \bibinfo {author} {\bibfnamefont {V.}~\bibnamefont
  {Rønning}}, \bibinfo {author} {\bibfnamefont {Y.}~\bibnamefont
  {Kalmann~Frodason}}, \bibinfo {author} {\bibfnamefont {V.}~\bibnamefont
  {Bobal}}, \bibinfo {author} {\bibfnamefont {L.}~\bibnamefont {Vines}}, \ and\
  \bibinfo {author} {\bibfnamefont {J.~B.}\ \bibnamefont {Varley}},\ }\href
  {\doibase 10.1103/PhysRevMaterials.4.074605} {\bibfield  {journal} {\bibinfo
  {journal} {Phys. Rev. Materials}\ }\textbf {\bibinfo {volume} {4}},\ \bibinfo
  {pages} {074605} (\bibinfo {year} {2020})}\BibitemShut {NoStop}%
\bibitem [{\citenamefont {Frodason}\ \emph {et~al.}(2021)\citenamefont
  {Frodason}, \citenamefont {Zimmermann}, \citenamefont {Verhoeven},
  \citenamefont {Weiser}, \citenamefont {Vines},\ and\ \citenamefont
  {Varley}}]{Frodason2021}%
  \BibitemOpen
  \bibfield  {author} {\bibinfo {author} {\bibfnamefont {Y.~K.}\ \bibnamefont
  {Frodason}}, \bibinfo {author} {\bibfnamefont {C.}~\bibnamefont
  {Zimmermann}}, \bibinfo {author} {\bibfnamefont {E.~F.}\ \bibnamefont
  {Verhoeven}}, \bibinfo {author} {\bibfnamefont {P.~M.}\ \bibnamefont
  {Weiser}}, \bibinfo {author} {\bibfnamefont {L.}~\bibnamefont {Vines}}, \
  and\ \bibinfo {author} {\bibfnamefont {J.~B.}\ \bibnamefont {Varley}},\
  }\href {\doibase 10.1103/PhysRevMaterials.5.025402} {\bibfield  {journal}
  {\bibinfo  {journal} {Phys. Rev. Materials}\ }\textbf {\bibinfo {volume}
  {5}},\ \bibinfo {pages} {025402} (\bibinfo {year} {2021})}\BibitemShut
  {NoStop}%
\bibitem [{\citenamefont {Korhonen}\ \emph {et~al.}(2015)\citenamefont
  {Korhonen}, \citenamefont {Tuomisto}, \citenamefont {Gogova}, \citenamefont
  {Wagner}, \citenamefont {Baldini}, \citenamefont {Galazka}, \citenamefont
  {Schewski},\ and\ \citenamefont {Albrecht}}]{Korhonen2015}%
  \BibitemOpen
  \bibfield  {author} {\bibinfo {author} {\bibfnamefont {E.}~\bibnamefont
  {Korhonen}}, \bibinfo {author} {\bibfnamefont {F.}~\bibnamefont {Tuomisto}},
  \bibinfo {author} {\bibfnamefont {D.}~\bibnamefont {Gogova}}, \bibinfo
  {author} {\bibfnamefont {G.}~\bibnamefont {Wagner}}, \bibinfo {author}
  {\bibfnamefont {M.}~\bibnamefont {Baldini}}, \bibinfo {author} {\bibfnamefont
  {Z.}~\bibnamefont {Galazka}}, \bibinfo {author} {\bibfnamefont
  {R.}~\bibnamefont {Schewski}}, \ and\ \bibinfo {author} {\bibfnamefont
  {M.}~\bibnamefont {Albrecht}},\ }\href {\doibase 10.1063/1.4922814}
  {\bibfield  {journal} {\bibinfo  {journal} {Applied Physics Letters}\
  }\textbf {\bibinfo {volume} {106}},\ \bibinfo {pages} {242103} (\bibinfo
  {year} {2015})}\BibitemShut {NoStop}%
\bibitem [{\citenamefont {Kresse}\ and\ \citenamefont
  {Hafner}(1993)}]{Kresse1993}%
  \BibitemOpen
  \bibfield  {author} {\bibinfo {author} {\bibfnamefont {G.}~\bibnamefont
  {Kresse}}\ and\ \bibinfo {author} {\bibfnamefont {J.}~\bibnamefont
  {Hafner}},\ }\href@noop {} {\bibfield  {journal} {\bibinfo  {journal} {Phys.
  Rev. B}\ }\textbf {\bibinfo {volume} {47}},\ \bibinfo {pages} {558} (\bibinfo
  {year} {1993})}\BibitemShut {NoStop}%
\bibitem [{\citenamefont {Kresse}\ and\ \citenamefont
  {Furthm{\"u}ller}(1996)}]{Kresse1996}%
  \BibitemOpen
  \bibfield  {author} {\bibinfo {author} {\bibfnamefont {G.}~\bibnamefont
  {Kresse}}\ and\ \bibinfo {author} {\bibfnamefont {J.}~\bibnamefont
  {Furthm{\"u}ller}},\ }\href@noop {} {\bibfield  {journal} {\bibinfo
  {journal} {Phys. Rev. B}\ }\textbf {\bibinfo {volume} {54}},\ \bibinfo
  {pages} {11169} (\bibinfo {year} {1996})}\BibitemShut {NoStop}%
\bibitem [{\citenamefont {Bl{\"o}chl}(1994)}]{Blochl1994}%
  \BibitemOpen
  \bibfield  {author} {\bibinfo {author} {\bibfnamefont {P.~E.}\ \bibnamefont
  {Bl{\"o}chl}},\ }\href@noop {} {\bibfield  {journal} {\bibinfo  {journal}
  {Phys. Rev. B}\ }\textbf {\bibinfo {volume} {50}},\ \bibinfo {pages} {17953}
  (\bibinfo {year} {1994})}\BibitemShut {NoStop}%
\bibitem [{\citenamefont {Heyd}, \citenamefont {Scuseria},\ and\ \citenamefont
  {Ernzerhof}(2003)}]{Heyd2003}%
  \BibitemOpen
  \bibfield  {author} {\bibinfo {author} {\bibfnamefont {J.}~\bibnamefont
  {Heyd}}, \bibinfo {author} {\bibfnamefont {G.~E.}\ \bibnamefont {Scuseria}},
  \ and\ \bibinfo {author} {\bibfnamefont {M.}~\bibnamefont {Ernzerhof}},\
  }\href {\doibase 10.1063/1.1564060} {\bibfield  {journal} {\bibinfo
  {journal} {J. Chem. Phys.}\ }\textbf {\bibinfo {volume} {118}},\ \bibinfo
  {pages} {8207} (\bibinfo {year} {2003})}\BibitemShut {NoStop}%
\bibitem [{\citenamefont {Heyd}, \citenamefont {Scuseria},\ and\ \citenamefont
  {Ernzerhof}(2006)}]{Heyd2006}%
  \BibitemOpen
  \bibfield  {author} {\bibinfo {author} {\bibfnamefont {J.}~\bibnamefont
  {Heyd}}, \bibinfo {author} {\bibfnamefont {G.~E.}\ \bibnamefont {Scuseria}},
  \ and\ \bibinfo {author} {\bibfnamefont {M.}~\bibnamefont {Ernzerhof}},\
  }\href {\doibase 10.1063/1.2204597} {\bibfield  {journal} {\bibinfo
  {journal} {J. Chem. Phys.}\ }\textbf {\bibinfo {volume} {124}},\ \bibinfo
  {pages} {219906} (\bibinfo {year} {2006})}\BibitemShut {NoStop}%
\bibitem [{\citenamefont {Freysoldt}\ \emph {et~al.}(2014)\citenamefont
  {Freysoldt}, \citenamefont {Grabowski}, \citenamefont {Hickel}, \citenamefont
  {Neugebauer}, \citenamefont {Kresse}, \citenamefont {Janotti},\ and\
  \citenamefont {{Van de Walle}}}]{Freysoldt2014}%
  \BibitemOpen
  \bibfield  {author} {\bibinfo {author} {\bibfnamefont {C.}~\bibnamefont
  {Freysoldt}}, \bibinfo {author} {\bibfnamefont {B.}~\bibnamefont
  {Grabowski}}, \bibinfo {author} {\bibfnamefont {T.}~\bibnamefont {Hickel}},
  \bibinfo {author} {\bibfnamefont {J.}~\bibnamefont {Neugebauer}}, \bibinfo
  {author} {\bibfnamefont {G.}~\bibnamefont {Kresse}}, \bibinfo {author}
  {\bibfnamefont {A.}~\bibnamefont {Janotti}}, \ and\ \bibinfo {author}
  {\bibfnamefont {C.~G.}\ \bibnamefont {{Van de Walle}}},\ }\href {\doibase
  10.1103/RevModPhys.86.253} {\bibfield  {journal} {\bibinfo  {journal} {Rev.
  Mod. Phys.}\ }\textbf {\bibinfo {volume} {86}},\ \bibinfo {pages} {253}
  (\bibinfo {year} {2014})}\BibitemShut {NoStop}%
\bibitem [{\citenamefont {Freysoldt}, \citenamefont {Neugebauer},\ and\
  \citenamefont {{Van de Walle}}(2011)}]{Freysoldt2011}%
  \BibitemOpen
  \bibfield  {author} {\bibinfo {author} {\bibfnamefont {C.}~\bibnamefont
  {Freysoldt}}, \bibinfo {author} {\bibfnamefont {J.}~\bibnamefont
  {Neugebauer}}, \ and\ \bibinfo {author} {\bibfnamefont {C.~G.}\ \bibnamefont
  {{Van de Walle}}},\ }\href {\doibase 10.1002/pssb.201046289} {\bibfield
  {journal} {\bibinfo  {journal} {Phys. Status Solidi B}\ }\textbf {\bibinfo
  {volume} {248}},\ \bibinfo {pages} {1067} (\bibinfo {year}
  {2011})}\BibitemShut {NoStop}%
\bibitem [{\citenamefont {Frodason}\ \emph {et~al.}(2023)\citenamefont
  {Frodason}, \citenamefont {Varley}, \citenamefont {Johansen}, \citenamefont
  {Vines},\ and\ \citenamefont {Van De~Walle}}]{Frodason2023}%
  \BibitemOpen
  \bibfield  {author} {\bibinfo {author} {\bibfnamefont {Y.~K.}\ \bibnamefont
  {Frodason}}, \bibinfo {author} {\bibfnamefont {J.~B.}\ \bibnamefont
  {Varley}}, \bibinfo {author} {\bibfnamefont {K.~M.~H.}\ \bibnamefont
  {Johansen}}, \bibinfo {author} {\bibfnamefont {L.}~\bibnamefont {Vines}}, \
  and\ \bibinfo {author} {\bibfnamefont {C.~G.}\ \bibnamefont {Van De~Walle}},\
  }\href {\doibase 10.1103/PhysRevB.107.024109} {\bibfield  {journal} {\bibinfo
   {journal} {Phys. Rev. B}\ }\textbf {\bibinfo {volume} {107}},\ \bibinfo
  {pages} {024109} (\bibinfo {year} {2023})}\BibitemShut {NoStop}%
\bibitem [{\citenamefont {Tuomisto}(2023)}]{Tuomisto2023}%
  \BibitemOpen
  \bibfield  {author} {\bibinfo {author} {\bibfnamefont {F.}~\bibnamefont
  {Tuomisto}},\ }\href {\doibase 10.35848/1347-4065/acc7b1} {\bibfield
  {journal} {\bibinfo  {journal} {Jpn. J. Appl. Phys.}\ }\textbf {\bibinfo
  {volume} {62}},\ \bibinfo {pages} {SF0802} (\bibinfo {year}
  {2023})}\BibitemShut {NoStop}%
\bibitem [{\citenamefont {Zhelezova}, \citenamefont {Makkonen},\ and\
  \citenamefont {Tuomisto}(2024)}]{Zhelezova2024}%
  \BibitemOpen
  \bibfield  {author} {\bibinfo {author} {\bibfnamefont {I.}~\bibnamefont
  {Zhelezova}}, \bibinfo {author} {\bibfnamefont {I.}~\bibnamefont {Makkonen}},
  \ and\ \bibinfo {author} {\bibfnamefont {F.}~\bibnamefont {Tuomisto}},\
  }\href {\doibase 10.1063/5.0205933} {\bibfield  {journal} {\bibinfo
  {journal} {Journal of Applied Physics}\ }\textbf {\bibinfo {volume} {136}},\
  \bibinfo {pages} {065702} (\bibinfo {year} {2024})}\BibitemShut {NoStop}%
\end{thebibliography}%

\beginsupplement

\newpage\section*{Supplementary Material}
\section{Chemical potentials}

The chemical potentials used to calculate the formation energies shown in Fig. 2 of the main text correspond to theoretical limiting cases. Under O-rich conditions, oxygen is assumed to be abundant in the growth environment, therefore $\mu_{\text{O}}=0$. Under O-poor conditions, the growth environment is assumed to be cation-rich, meaning that in Ga$_2$O$_3$, $\mu_{\text{Ga}}=0$. The non-zero chemical potential is then calculated by satisfying the stability condition  

\begin{align}\label{eq:ga2o3_enthalpy} 
&\Delta\text{H}(\text{Ga}_{2}\text{O}_{3}) = 2\mu_{\text{Ga}} + 3\mu_{\text{O}}. 
\end{align}

However, when considering defects in an alloy such as AlGaO$_3$, the choice of these chemical potentials is ambiguous. Here, for O-poor conditions, $\mu_{\text{Ga}}$=0 and $\mu_{\text{O}}$ are calculated using AlGaO$_3$ as a reference, from the following

\begin{align}\label{eq:algao_enthalpy} 
&\Delta\text{H}(\text{AlGaO}_{3}) = \mu_{\text{Al}} + \mu_{\text{Ga}} + 3\mu_{\text{O}}. 
\end{align}

For extrinsic dopants, the chemical potential for the dopant must be calculated using a limiting phase. For Si, $\mu_{\text{Si}}$, was calculated using SiO$_2$ as a reference to satisfy

\begin{align}\label{eq:sio2_enthalpy} 
&\Delta\text{H}(\text{SiO}_{2}) = \mu_{\text{Si}} + 2\mu_{\text{O}}. 
\end{align}

\begin{table}[b]
	\caption{\label{tab:chem_pots} Chemical potentials (in eV) for O-rich and O-poor conditions for all relevant species in Al$_2$O$_3$, AlGaO$_3$ and Ga$_2$O$_3$ }
ALT TEXT:  In O-rich conditions, the oxygen chemical potential is always zero. The Al and Ga chemical potentials are In O-poor conditions the chemical potentials for the cations in aluminum oxide and gallium oxide are zero, and the Ga chemical potential in the 50-50 alloy is set to zero. 
	\begin{ruledtabular}
	\begin{center}
	\centering
		\begin{tabular}{llll}
 	         	& Al$_2$O$_3$& AlGaO$_3$&Ga$_2$O$_3$\\
			\hline
			O-rich & & &\\
			\hline
			$\mu_{\text{Al}}$ & $-$8.124& $-$8.124&-\\
			$\mu_{\text{Ga}}$  &- & $-$5.191&$-$5.191\\
			$\mu_{\text{O}}$ &0& 0& 0 \\
			$\mu_{\text{Si}}$  & $-$9.090& $-$9.090& $-$9.090\\
			\hline
			O-poor & & &\\
			\hline
			$\mu_{\text{Al}}$ & 0 & $-$2.934&-\\
			$\mu_{\text{Ga}}$ &- & 0&0\\
			$\mu_{\text{O}}$ & $-$5.416& $-$3.460&$-$3.463\\
			$\mu_{\text{Si}}$ &0& $-$2.169& $-$2.162
		\end{tabular}
\end{center}
\end{ruledtabular}
\end{table}

Table~\ref{tab:chem_pots} lists the chemical potentials used to calculate the formation energies for the cation vacancies and Si in Al$_2$O$_3$, AlGaO$_3$ and Ga$_2$O$_3$.

\end{document}